\documentclass[10pt,superscriptaddress,twocolumn,amsmath,amssymb,fleqn]{revtex4}
\usepackage{graphicx}
\usepackage{dcolumn}
\usepackage{bm}

\begin{document}

\title{Evidence of breakdown of the spin symmetry in diluted 2D electron gases}
\author{P. Giudici}
\affiliation{Institut f\"ur Festk\"orperphysik, Technische
Universit\"at Berlin, Hardenbergstr. 36, 10623 Berlin, Germany}

\author{A. R. Go\~ni}
\affiliation{ICREA Research Professor, Institut de Ci\`encia de Materials de Barcelona, Campus de la UAB, 08193 Bellaterra, Spain}

\author{P. G. Bolcatto}
\affiliation{Facultad de Ingenier\'{\i}a Qu\'{\i}mica and Facultad de Humanidades y Ciencias,
Universidad Nacional del Litoral, 3000 Santa Fe, Argentina}
\author{C. R. Proetto}
\affiliation{Centro At\'omico Bariloche and Instituto Balseiro, 8400 S. C. de
Bariloche, R\'{\i}o Negro, Argentina}

\author{K. Eberl}
\affiliation{MPI f\"ur Festk\"orperforschung, Heisenbergstr. 1,
70569 Stuttgart, Germany}

\author{M. Hauser}
\affiliation{MPI f\"ur Festk\"orperforschung, Heisenbergstr. 1,
70569 Stuttgart, Germany}

\author{C. Thomsen}
\affiliation{Institut f\"ur Festk\"orperphysik, Technische
Universit\"at Berlin, Hardenbergstr. 36, 10623 Berlin, Germany}

\date{\today}
\begin{abstract}
\end{abstract}

\maketitle


Recent  claims of an experimental demonstration of spontaneous
spin polarisation in dilute electron gases~\cite{young99} revived
long standing theoretical discussions~\cite{ceper99,bloch}. In two
dimensions, the stabilisation of a ferromagnetic fluid might be
hindered by the occurrence of the metal-insulator transition at
low densities~\cite{abra79}. To circumvent localisation in the
two-dimensional electron gas (2DEG) we investigated the low
populated second electron subband, where the disorder potential is
mainly screened by the high density of the first subband. This
letter reports on the breakdown of the spin symmetry in a 2DEG,
revealed by the abrupt enhancement of the exchange and correlation
terms of the Coulomb interaction, as determined from the energies
of the collective charge and spin excitations. Inelastic light
scattering experiments and calculations within the time-dependent
local spin-density approximation give strong evidence for the
existence of a ferromagnetic ground state in the diluted regime.

Recently we have shown  experimentally as well as theoretically
that at low temperatures and zero magnetic field a 2D electron gas
realized in a GaAs single quantum well (SQW) undergoes a
first-order phase transition, as result of  intersubband
exchange-correlation terms when the second electron subband
becomes populated \cite{goni04}. The situation of two occupied
subbands resembles that of double layer structures, a system with
a rich ground state phase diagram, where  a collapse of the normal
metallic state towards a spontaneous magnetic order at very low
densities was also predicted~\cite{reb97,bolc03}. Evidence of this first
order phase transition is the abrupt renormalisation of the second
subband energy and its sudden population with
electrons~\cite{goni02} as observed by photoluminescence (PL)
spectra. The first-order character of the transition was confirmed
by self-consistent density-functional calculations, including
exchange interactions exactly~\cite{rebor03,rigam03}. The exact-exchange theory
predicted also a spontaneous breakdown of the spin symmetry to
take place for a low population of the second subband.  In this
situation a new phase was found in PL spectra at low temperatures
and at zero magnetic field~\cite{goni04}. The behaviour of this
phase by varying density, temperature and external magnetic field
speaks for the formation of spin-polarised domains with different
in-plane magnetisation, as expected for planar ferromagnets to
minimise the stray field. Moreover, last year possible ferromagnetism was
claimed to be observed in transport experiments on a GaAs heterostructure with a single occupied
subband~\cite{ghosh04}. This evidence, though, is not conclusive due to the
effects of localisation mentioned above. In this work we
investigate the elementary excitations of the 2DEG, observing a
strong enhancement of the exchange-correlation contributions
occurring at very low population of the second subband.
Density-functional calculations within the time-dependent local
spin-density approximation (TDLSDA) confirm the existence of such
spin polarised phase.


The 2DEG forms in modulation-doped GaAs/Al$_{0.33}$Ga$_{0.67}$As
single quantum wells of 20, 25 and 30 nm well thickness grown by
molecular-beam epitaxy. The 2DEG has a mean mobility of about
$8\times 10^5\,$cm$^2\,$/Vs and a  density such that only the
lowest subband is occupied, with Fermi energies $E_{F}$ between 25
and 28 meV at 4.2$\,$K. The electron density can be changed by
applying a dc bias between the 2DEG and a metallic back contact.
Photoluminescence and inelastic light scattering spectra were
obtained using low power densities in the range of 0.5 to 5
W/cm$^2$. Light scattering spectra were recorded in backscattering
geometry using a line focus in order to avoid heating of the 2DEG
at the required laser powers.

\begin{figure}[t]
\par
\begin{center}
\leavevmode
\includegraphics[width=1.0\linewidth]{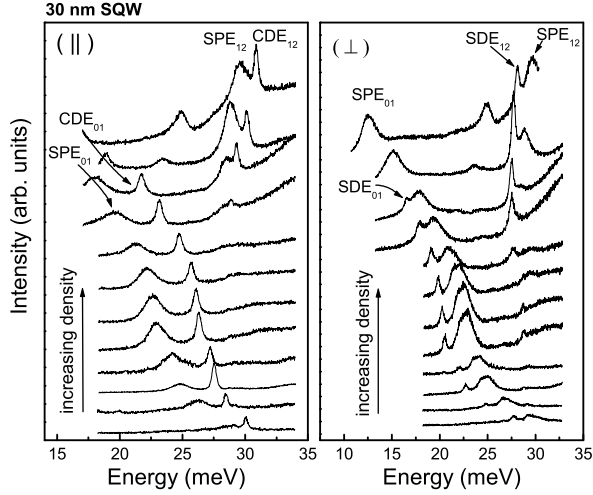}
\end{center}
\caption{Inelastic light scattering spectra of a 30 nm  wide SQW
obtained at  2$\,$K, with different voltages and in polarised
($\parallel$) and depolarised ($\perp$) configuration. In the
spectra with parallel polarisation the CDE associated with
intersubband transitions between the first and the second subband
$ 0 \to 1 $, and between the second and the third subbands $ 1 \to
2 $ are apparent. The corresponding SDE, in contrast, are observed
in depolarised configuration. The broader peak that appears in
spectra with both polarisations is assigned to the single-particle
excitation SPE and is centred at the intersubband energy $E_{ij}$,
with $(ij)$ being $(01)$ or $(12)$.} \label{fig:esp}
\end{figure}

The set of spectra in Fig.~\ref{fig:esp}  illustrates the
evolution of the energy of the charge (CDE), spin (SDE) and
single-particle excitations (SPE), associated with the
intersubband transitions $ 0 \to 1 $ and  $ 1 \to 2 $, when
varying the density of the electron gas.  The shift of the
SPE$_{01}$ to lower energies and of the SPE$_{12}$ to higher
energies are consequences of the renormalisation of the second
subband with increasing occupation. Furthermore, the change in the
energy separation between excitations indicates a dependence of
the many-body corrections on electron
density~\cite{pincz89a,gammon90}.

As explained in the Appendix, within the formalism of
TDLSDA~\cite{kohn} and in the long wavelength limit ($q\approx 0$
for backscattering geometry), the energies of the charge and
spin-density excitation $ \hbar\omega_{\mu}$ can be written in the
following compact way~\cite{bolc03}
\begin{eqnarray} \label{eq:uno}
(\hbar\omega_{\mu})^2 - E_{i,i+1}^2 &=& 2~\delta{n}_{i,i+1}
~E_{i,i+1}\text{ }\gamma^{i,i+1}_{\mu},
\end{eqnarray}
\noindent with $\delta{n}_{i,i+1}= n_{i}-n_{i+1}$ being the
difference in the subband densities and $E_{i,i+1}=E_{i+1}-E_{i}$,
where $E_i$ are the subband energies. In the paramagnetic (P)
phase, the index $\mu=\,$CDE,$ \, $SDE describes the uncoupled
charge and spin density excitations. According to standard
notations,
$\gamma^{i,i+1}_{CDE}=(\alpha^{i,i+1}/\tilde{\epsilon}-\beta^{i,i+1}_{CDE})$,
and $\gamma^{i,i+1}_{SDE}=-\beta^{i,i+1}_{SDE}$.
 $\tilde{\epsilon}(\omega)$ is the contribution of the polar
lattice to the dielectric function~\cite{pincz89a}. In the ferromagnetic phase (F), charge and spin are
intrinsically linked, thus, there is a single collective mode of
mixed character (denoted with subindex $\mu=\,$M). The
depolarisation shift $\alpha$ is a positive Hartree contribution
that in the paramagnetic phase typically dominates over the exchange-correlation term
$\beta_{CDE}$, whereas $\gamma$ is dominated by $\beta$ in the ferromagnetic case. The explicit
expressions for $\alpha, \beta_{CDE}, \beta_{SDE}$, and
$\gamma_{M}$ in terms of matrix elements of the Hartree and
exchange-correlation potentials are given in the Appendix.

\begin{figure}[h]
\par
\begin{center}
\leavevmode
\includegraphics[width=1.0\linewidth]{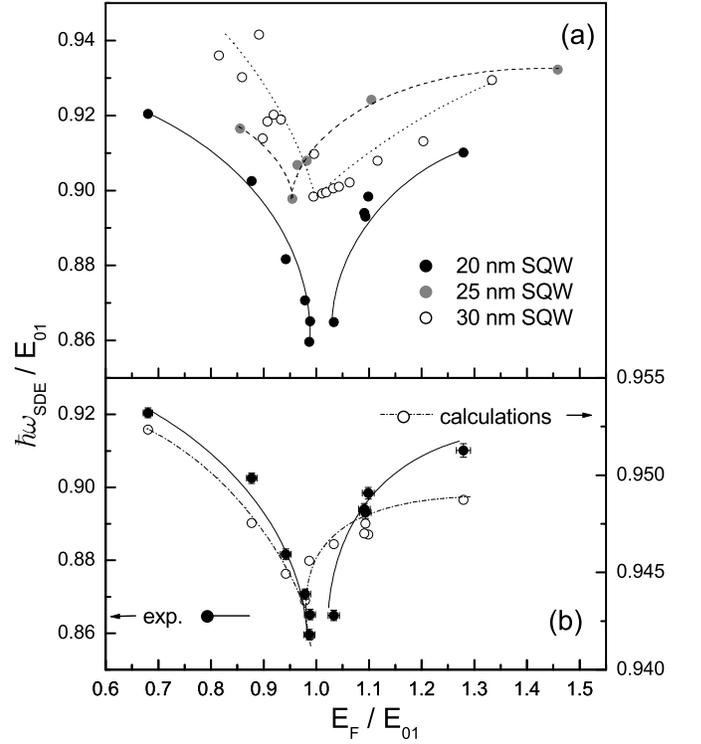}
\end{center}
\caption{The $ 0 \to 1 $ spin-density mode softening as a function
of the dimensionless parameter $E_{F}$/$E_{01}$, being
$E_{F}$/$E_{01}=1$ the borderline between one and two occupied
subbands. a) Comparison of the SDE softening among the 20, 25, and
30 nm wide SQW. b) Comparison of the experimental SDE softening
(solid circles, left axis) for the 20 nm wide SQW with TDLSDA
results (open circles, right axis). The lines are guides to the
eye.} \label{fig:sdex}
\end{figure}

In Fig.~\ref{fig:sdex}  we show an unexpected behaviour of the
spin-density mode: it becomes softer, i.e. decreases in absolute
value when the Fermi level is resonant with the second subband,
recovering the normal behaviour at higher densities. From the
study of the formation of a non trivial magnetic order,
intensively investigated in double-layer systems~\cite{bolc00}, a
softening of the SDE$_{01}$ can be considered as an indication of
ferromagnetism in the second subband. In our system the softening
is rather weak, probably due to the large contribution of
electrons in the paramagnetic first subband. Comparing the three
QW widths (Fig.~\ref{fig:sdex}(a)) it can be seen that the SDE
softening is enhanced for the narrowest QW. That speaks sensibly
for a dependence of the possible ferromagnetic order on the
strength of the wave function confinement. In
Fig.~\ref{fig:sdex}(b) we compare the experimental data (solid
points) with the excitation energies yielded by TDLSDA
calculations (empty points). The left (right) axis corresponds to
the experimental (calculated) results. The underestimation of the
softening resulting from the calculations can be attributed to
shortcomings of the TDLSDA in the treatment of exchange and
correlation but also to the fact that the calculations always
yield a paramagnetic ground state for this high total density.

\begin{figure}[h]
\par
\begin{center}
\leavevmode
\includegraphics[width=1.0\linewidth]{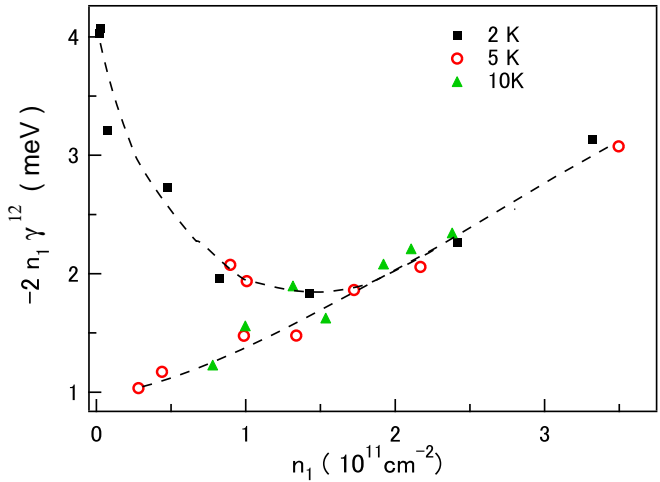}
\end{center}
\caption{Enhancement of the exchange-correlation energy
contribution $2 n_1\gamma^{12}$, as a function of the density
$n_1$ of the second subband, for the 30$\,$nm wide SQW and at
temperatures T$=2\,$K (circles), T$=5\,$K (squares) and T$=10\,$K
(triangles). Curves are a guide to the eye. } \label{fig:beta1}
\end{figure}

For this reason and in order to exclude the effect of the
electrons in the first paramagnetic subband, we investigate higher
intersubband excitations associated with the transition $1\to 2$.
These excitations occur between the sparsely populated second
subband and the empty third  subband. Only electrons from the spin
polarised phase in $E_{1}$ contribute to the SDE$_{12}$. In the
series of spectra of Fig.~\ref{fig:esp} the excitations
SPE$_{12}$, SDE$_{12}$ and CDE$_{12}$ become apparent when the
second subband is populated. In particular, the CDE$_{12}$ shows
up exactly on top of the SPE$_{12}$ due to the collapse of the
Hartree term $\alpha^{12}$ at very low densities~\cite{ernst94}.
In contrast, the SDE$_{12}$ is always shifted down from the
SPE$_{12}$. From the measured energies of the excitations
obtained with cross polarisation and using Eq.~(\ref{eq:uno}) we
determine the many-body correction energy ($2
n_1\gamma^{12}$). Multiplying by $n_1$ we avoid divergencies
when the density goes to zero. The results are shown in
Fig.~\ref{fig:beta1}.
The striking result of this work is the observation of an
enhancement of the exchange-correlation contribution at low
temperature when the system enters the region of the
spin-instability. This leads to a pronounced softening  of the
collective mode associated with electronic transitions $1 \to 2$.
The diluted electron gas of the second subband becomes thus
unstable against spin-flip excitations, which trigger the
transition into the ferromagnetic phase. By increasing the
temperature a scaling down of the exchange-correlation energy
occurs, in accordance with the disappearance of the new phase
observed in the PL spectra, as discussed in previous work
\cite{goni04}.

For a better understanding of the physical origin  of this
transition it is instructive to perform the corresponding
calculation of the excitations. To avoid the effect of the high
density in the first subband we constructed an auxiliary
structure, using for its intersubband energy $E_{01}$ and density
$n_{0}$, the experimental values $E_{12}$ and $n_{1}$,
respectively. Doing so, the calculations can yield now a
ferromagnetic ground state and the corresponding excitations.
\begin{figure}[h]
\par
\begin{center}
\leavevmode
\includegraphics[width=\linewidth]{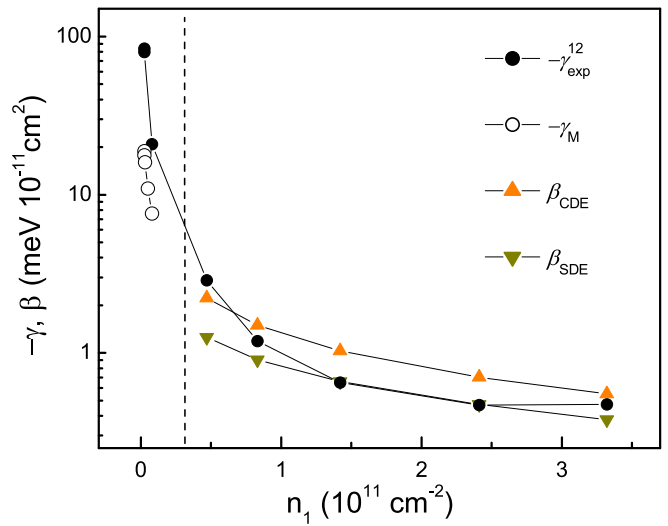}
\end{center}
\caption{Comparison between the calculated many-body corrections
$\gamma_{M}$ (ferromagnetic solution), $\beta_{CDE}$ and
$\beta_{SDE}$ (paramagnetic solution), obtained within TDLSDA and
$\gamma^{12}_{exp}$ from depolarised spectra at 2$\,$K of the 30 nm
wide SQW. The dashed line divides the ferromagnetic (left) from
the paramagnetic (right) region.} \label{fig:beta2}
\end{figure}

Figure 4 compares the experimental $\gamma^{12}_{exp}$ values, as
obtained from the excitations measured in cross polarisation at 2
K, with the calculated ones. TDLSDA predicts the occurrence of a
ferromagnetic phase, characterised by $\gamma_{M}$ (open circles),
for densities below 0.34$\times 10^{11}$cm$^{-2}$ and a
paramagnetic phase otherwise, characterised by the excitonic
shifts $\beta_{CDE}$ and $\beta_{SDE}$ (triangles). Inspection of
the different terms in $\gamma_{M}$ reveals that its strong
increase at low densities is mainly due to the fact that the
contribution from correlation has the opposite sign than the one
to $\beta_{SDE}$ in the paramagnetic phase. Thus, correlation adds
to the effect of exchange in the spin-polarised case, which favours
ferromagnetism.  In fact, the calculated values of $\beta_{SDE}$
and $\gamma_{M}$ at high and low densities, respectively, agree
well with $\gamma^{12}_{exp}$. The range of stability of the
ferromagnetic solution is also in nice agreement with the data of
Fig. 3, where the strong enhancement of exchange-correlation vertex corrections,
observed at low temperatures, occurs at densities below $1\times
10^{11}$cm$^{-2}$. In this way, we have obtained from light
scattering experiments and local spin-density calculations the
first compelling evidence for the existence of a ferromagnetic
ground state of 2DEGs in an effectively diluted regime; one of the
exciting puzzles of many-body physics stated more than fifty years
ago by F. Bloch \cite{bloch}.

\appendix
\section{Methods}

The ground-state calculations have been performed in the local
spin-density approximation (LSDA) in the framework of
density-functional theory. We adjusted the geometry of the SQW in
order to reproduce the Fermi energy $E_{F}$, and the two subbands
energies $E_{0}$ and $E_{1}$ as obtained from photoluminescence
experiments.

The energy of the collective excitations of charge (CDE) and spin
(SDE) were obtained with TDLSDA from the zeros of the secular
determinant of the spin-dependent dielectric tensor $(\det\epsilon
= 0)$, which can be written in terms of the Hartree and
exchange-correlation potentials in the subband representation
\cite{bolc03}. The corresponding potential matrix elements, as a
function of the wavevector $q$ of the excitations, are

\begin{eqnarray}
\label{eq:potencial1} \left(V_{H}\right)_{ij,i^{\prime }j^{\prime
}}^{\sigma ,\sigma ^{\prime }}(q) &=&\frac{1}{A} \int dz\int
dz^{\prime } (2\pi e^{2}/\varepsilon q)\exp (-q\left| z-z^{\prime
}\right| )
\nonumber \\
&\times &\phi _{i\sigma }^{*}\left( z\right) \phi _{j\sigma
}\left( z\right) \phi _{i^{\prime }\sigma ^{\prime }}^{*}\left(
z^{\prime }\right) \phi _{j^{\prime }\sigma ^{\prime }}\left(
z^{\prime }\right),
\end{eqnarray}

\begin{eqnarray}
\label{eq:potencial2} \left(V_{xc}\right)_{ij,i^{\prime }j^{\prime
}}^{\sigma ,\sigma ^{\prime }}(q) &=&\frac{1}{A} \int dz\phi
_{i\sigma }^{*}\left( z\right) \phi _{j\sigma }\left( z\right)
\phi _{i^{\prime }\sigma ^{\prime }}^{*}\left( z\right) \phi
_{j^{\prime }\sigma ^{\prime }}\left( z\right)
\nonumber \\
&\times & \left[ K_{xc}(z)+(\sigma +\sigma ^{\prime
})J_{xc}(z)+\sigma \sigma ^{\prime }I_{xc}(z)\right]. \nonumber \\
&
\end{eqnarray}

\noindent $\phi _{i\sigma }\left( z\right) $ are the
self-consistent wave functions that diagonalise the effective
one-dimensional LSDA\ Hamiltonian, after assuming translational
invariance along the $(x,y)$ plane (area $A$). $i,j,i^{^{\prime
}},j^{\prime }$ are subband indexes, and $\sigma,
\sigma^{\prime}=1(-1)$ for spin $\uparrow (\downarrow)$. Eq.
(\ref{eq:potencial1}) corresponds to the Hartree contribution,
with $\varepsilon$ being the dielectric constant of GaAs. Eq.
(\ref{eq:potencial2}) corresponds to the exchange-correlation
contribution, with $K, J,$ and $I$ being derivatives of the
exchange-correlation energy functional with respect to the density
and magnetisation of the 2DEG \cite{bolc03}. Making a two-subband
approximation $(i=i^{\prime}, j=j^{\prime})$, and in the long
wavelength limit $(q\rightarrow 0)$, the eigenmode equation
$\det\epsilon = 0$ can be solved analytically for
$(\hbar\omega)^{2}$. The resulting equations for the P phase take
the form of Eq.~(\ref{eq:uno}), provided the following
identification is made: $\alpha=V_{H},
\beta_{CDE}=-\left(V_{xc}^{\uparrow\uparrow}+V_{xc}^{\uparrow\downarrow}\right)/2$,
and
$\beta_{SDE}=-\left(V_{xc}^{\uparrow\uparrow}-V_{xc}^{\uparrow\downarrow}\right)/2$.
For the sake of simplicity, we have eliminated subband indexes,
and we have used the symmetry
$V_{H}=V_{H}^{\uparrow\uparrow}=V_{H}^{\downarrow\downarrow}=V_{H}^{\uparrow\downarrow}=V_{H}^{\downarrow\uparrow}$.
In the F phase,
$\gamma_{M}=(V_{H}^{\uparrow\uparrow}+V_{xc}^{\uparrow\uparrow})$.
This {\it single} collective mode has a mixed CDE and SDE
character, and accordingly it has both Hartree and
exchange-correlation contributions, the latter being the dominant term in the F phase.

\begin{acknowledgments} We thank W. Dietsche for  assistance with the sample design and growth. P. G. acknowledges financial support from the DAAD and the ``Berliner Programm zur F\"orderung der Chancengleichheit''. This work is supported in part by the DFG in the framework of Sfb 296. P. G. B. acknowledges to U.N.L., ANPCyT and Fundaci\'{o}n Antorchas by Grants PE-213, 03-11736 and 14248-128 respectively. C. R. P. and P. G. B. are partially supported by CONICET.
\end{acknowledgments}

\end{document}